\documentclass[10pt,journal,final,twocolumn,]{IEEEtran}

\ifCLASSINFOpdf

\else

\fi

\usepackage{amsmath}
\usepackage{graphicx}

\begin{document}

\title{Three-dimensional compacted optical waveguide couplers designed by invariant engineerings}
\author{{Hongying Liu and L. F. Wei}
	\thanks{Hongying Liu is Information Quantum Technology Laboratory, International Cooperation Research Center of China Communication and Sensor Networks for Modern Transportation, School of Information Science and Technology, Southwest Jiaotong
University, Chengdu 610031, China. (e-mail: 893315258@qq.com)}
\thanks{L. F. Wei is Information Quantum Technology Laboratory, International Cooperation Research Center of China Communication and Sensor Networks for Modern Transportation, School of Information Science and Technology, Southwest Jiaotong
University, Chengdu 610031, China (e-mail: lfwei@swjtu.edu.cn)}

\thanks{L. F. Wei is also Information Quantum Technology Laboratory and Institute of Functional Materials,
College of Science, Donghua University, Shanghai 01620, China (email: lfwei@dhu.edu.cn)}}

\maketitle

\begin{abstract}
Due to the limitations either on the sizes of devices and signal routing channels, the current planar integrated optical waveguide circuits await for the further developments into the three-dimensional (3D) integrations, although their designs and fabrications are still challenges. In this paper we demonstrate an analytical method, basing on the invariant engineering, to overcome the complication in the usual method by numerically solving the relevant 3D coupled-mode equations for designing various 3D optical waveguide devices such as the typical couplers. Our method is based on the quantum-optical analogy, i.e., the Maxwell equation for the electrcomagnetic wave prorogating along the waveguide structure in the spatial domain is formally similar to the Schr\"odinger equation for the evolving quantum state in the time domain. We find that the spatial-domain invariants can be effectively constructed to solve the 3D coupled-mode equations, analogously to solve the dynamical evolutions of quantum systems in the time-domain. As a consequence, as long as appropriately set the coupling parameters between the 3D interconnected waveguides, the 3D three-waveguide couplers could be designed for various desirably power divisions. As the invariant method is a natural shortcut to the adiabaticity, the compacted devices designed by the invariant-based engineerings are robust against the coupling coefficient variations and the coupler lengths.
\end{abstract}


\IEEEpeerreviewmaketitle

\section{Introduction}

\IEEEPARstart{I}{ntegrated} optics technique has been rapidly developed to meet the growing demands in optical communications. Currently, the most conventional integrated optical devices are fabricated on the two-dimensional planes~\cite{book}.
That is to say, it can only process optical signals in a single plane space, and thus the number of data channels on an integrated chip is limited by its horizontal dimension. In order to high effectively implement the desired large data streams, image processing, neural networks, optical parallel logic operations and optical interconnection of integrated circuit chips, etc., the data transmissions between different layers to expand the more data channels are required naturally. Therefore, designing and fabricating various three-dimensional (3D) integrated optical devicesis are particularly expectable~\cite{WC1,GS,NS,SG,SB}. Naturally, the 3D integrations provide the desired additional dimensions and thus can effectively expand the number of communication channels. In practical, based on the reliability and structural stability of the current two-dimensional (2D) integrated optical technique, the 3D integrated optical technology for implementing the 3D routings is feasible~\cite{book}.

Traditionally, the designs of waveguide devices are based on the solutions of the relevant coupled-mode equations, which are delivered from the corresponding Maxwell equations describing the transports of electromagnetic waves in the waveguide structures. In the coupled waveguides, the coupled-mode equations are usually too complicate to be analytically solved and thus designing the devices by numerical methods to realize various desirable functions are very inconvenient. Recently, using the quantum-optical analogies to simplify the solutions of the coupled-mode equations has being attracted much attention in recent years~\cite{SL,HY,SY1,SY2,KH,EP,AS,KC}. This is because that the Maxwell equation for the electromagnetic wave prorogating along the curved waveguide structure in the spatial domain is formally similar to the Schr\"odinger equation for the quantum state evolving in the time domain.

Given various analytical methods to solve the Schr\"odinger equations have been developed well for the evolutions of the relevant quantum systems, various quantum-optical analogies are useful to solve the coupled-mode equations for designing the various waveguide devices. For example, the adiabatic coherent quantum algorithms, typically such as the atomic stimulated Raman adiabatic passages~\cite{KB} and coherent tunnelling adiabatic passage~\cite{AD}, for coherently manipulating the population transfers between the selected quantum states, have been used analogously to design certain waveguide devices wherein the intensity distributions of the light could be controlled deterministically. Specifically, the planar directional optical couplers had been designed by applying the analogies of the adiabatic quantum algorithms~\cite{EP,AS,KC}, and sequentially can be fabricated and verified experimentally~\cite{MM}. One of the advantages of the adiabatic designs is that, the coupling efficiency is insensitivity against the settings of the inter-waveguide coupling parameters and also the wavelengths of the waveguides. However, due to the limitation of adiabatic criterion, the lengths of the coupled regimes can not be shortened easily. As a consequence, the method of shortcuts to the adiabaticity is necessary~\cite{HY,HY2} for lowering the sizes of the designed devices.

Beyond the adiabatic limit, several adiabatic shortcut methods to implement the fast population transfers in two- and three-level atomic systems have been demonstrated~\cite{DM,BM,CX1,MS,CX2,BMG,IS2,FM} in quantum physics. Basically, to control the  fast evolution of a driven quantum system, we need to exactly solve the time-dependent Schr\"odinger equation
\begin{equation}
j\hbar\frac{\partial}{\partial t}|\Psi(t)\rangle=H(t)|\Psi(t)\rangle,
\end{equation}
with the parameter-designable Hamiltonian $H(t)$. However, such a solution is not easy to be obtain without the adiabatic approximation. Fortunately, if a so-called dynamical invariant $I(t)$ satisfying the condition:
\begin{equation}
\frac{dI(t)}{dt}=\frac{\partial I(t)}{\partial t}-\frac{1}{j\hbar}[H(t),I(t)]=0,
\end{equation}
can be found, then the generic solution to the corresponding Schr\"odinger equation (1) can be constructed as
\begin{equation}
|\Psi(t)\rangle=\sum_{n}c_{n}e^{j\alpha_{n}(t)}|\Phi_{n}(t)\rangle.
\end{equation}
Here, $|\Phi_{n}(t)\rangle$ is the instantaneous eigenstate of the dynamical invariant $I(t)$ corresponding to the time-independent eigenvalue $\lambda_n$, and  $\alpha_{n}(t)=\hbar^{-1}\int_{0}^{t}\langle\Phi_{n}(t^{'})|j\hbar\partial/\partial t'-H(t')|\Phi_{n}(t')\rangle dt'$
the Lewis-Riesenfeld phase~\cite{LH}. With such an exact solution, the quantum evolution of the driven system can be engineered by designing the proper time-dependent Hamiltonian. Obviously, the invariant method provides an effective approach to implement the fast quantum controls~\cite{LH,MJ,CX3,CX4,SD,SJ,SJ1}, and could be applied to design the compacted optical waveguide devices by using the quantum-optical analogies~\cite{SY3}.

In this paper, we develop a spatial-domain invariant method to exactly solve the coupled-mode equations for the three-waveguide structures and discuss how to engineer the inter-waveguide coupling parameters for designing the relevant 3D integrated optical waveguide devices. Certainly, this is a natural generalization of the shortcuts to adiabaticity developed previously for designing the compacted waveguide devices~\cite{HY2,KH,SY3} in a plane. Probably, one of the most advantages in this method is that the coupled-mode equations, for the three-dimensional waveguide structures~\cite{FD,WC2} with various inter-waveguide coupling parameters, can be solved analytically, rather than numerically~\cite{AC,AB}. This provides an effective approach to design the compacted 3D waveguide devices specifically, e.g., the couplers for realizing designable power distributions between the waveguide in different layers. Hopefully, our approach provides an attractive solution to design various ultra-dense 3D integrated optical devices and also demonstrate novel quantum-optical effects.

\begin{figure}[!t]
	\centering
	\includegraphics[width=8.5cm,height=4.5cm]{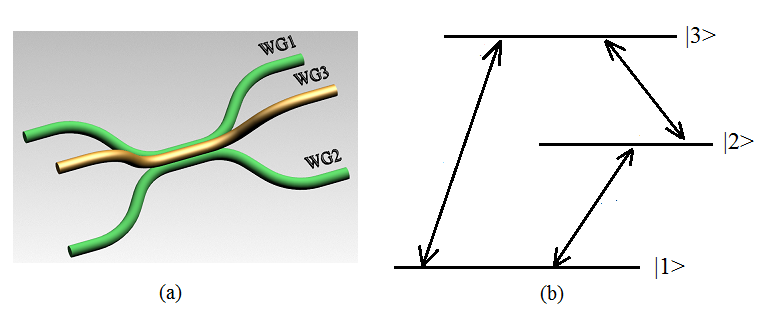}
		\caption{(a) A three-dimensional three-waveguide coupler, which is corresponding to a driven $\Delta$-type three-level quantum system (b).}
\label{fig_1}
\end{figure}

\section{Exactly solving the coupled-mode equations for three-waveguide structures with arbitrary inter-waveguide couplings by invariant method}

 Let us consider a relatively simple three-waveguide structure shown schematically in Fig.~1(a), i.e., a 3D waveguide coupler. Here, each waveguide can be coupled to the other two waveguides. For the simplicity, we suppose that one waveguide supports only one mode, i.e., the waveguide $m~(m=1,2,3)$ supports the propagating mode $|\Psi_m\rangle$, with
 \begin{align}
|\Psi_1\rangle=\begin{bmatrix} 1\\0\\0\end{bmatrix},\,
|\Psi_2\rangle=\begin{bmatrix} 0\\1\\0\end{bmatrix},\,
|\Psi_3\rangle=\begin{bmatrix} 0\\0\\1\end{bmatrix}.
\end{align}
The electric field of the light at the arbitrary spatial point in the coupler can be generically expressed as $E(x,y,z)=\sum_{m=1,2,3} a_{m}(z)exp(-j\beta_{m}z)|\Psi_{m}\rangle$ with $a_{m}(z)$ and $\beta_{m}$ being the amplitude and propagation constant of the $m$th mode at the spatial point $z$ along the propagation direction, respectively.

Based on the Maxwell equations, the propagation of the light in the present coupled-waveguide system can be expressed by the following coupled-mode equation~\cite{EPe}:
\begin{align}
j\frac{d}{dz}a_{m}(z)=\sum_{m\neq l=1,2,3}\Omega_{ml}(z)a_{l}(z)e^{-j\delta\beta_{m,l}},
\end{align}
with $\Omega_{ml}$ being the coupling coefficient between the $m$-th and the $l$-th waveguides, and $\delta\beta_{m,l}=\beta_m-\beta_l$. Note that the coupling coefficient $\Omega_{ml}$ could be complex for the multilayer waveguide couplings~\cite{KHS,ENE}.
Let
$A_1(z)=a_{1}(z)e^{-j\delta\beta_{1,2}z}, A_2(z)=a_{2}(z)$, and $A_3(z)=a_{3}(z)e^{-j\delta\beta_{3,2}z}$, we have
\begin{align}
j\frac{d}{dz}
\begin{bmatrix} A_{1}\\A_{2}\\A_{3}\end{bmatrix}&=
\begin{bmatrix} \delta\beta_{1,2}&\Omega_{12}&\Omega_{13}
\\\Omega_{21}&0&\Omega_{23}\\\Omega_{31}&\Omega_{32}
&\delta\beta_{3,2}\end{bmatrix}
\begin{bmatrix}A_{1}\\A_{2}\\A_{3}
\end{bmatrix}.
\end{align}

Specifically, when $\delta\beta_{1,2}=\delta\beta_{3,2}=0$, the coupled-mode equation reduces to
 \begin{align}
j\frac{d}{dz}\begin{bmatrix} A_{1}\\A_{2}\\A_{3}\end{bmatrix}&=H(z)
\begin{bmatrix}A_{1}\\A_{2}\\A_{3}\end{bmatrix},
\end{align}
with
\begin{align}
H(z)&=\begin{bmatrix} 0&\Omega_{12}(z)&\Omega_{13}(z)\\ \Omega_{21}(z)&0&\Omega_{23}(z)\\\Omega_{31}(z)&\Omega_{32}(z)&0\end{bmatrix}.
\end{align}
This equation is formally equivalent to the time-dependent Schr\"odinger equation for a resonantly-driven $\Delta$-type three-level artifical quantum system with close-loop dipole-transition structure~\cite{YXLIU,WZJ}, see, Fig.~1(b), by just replacing the time variable $t$ as the spatial variable $z$. Therefore, the present problem for the light power transfers between the coupled waveguides is analogous to that for the population transfers between the three atomic levels. The roles of the designable inter-waveguide coupling coefficients in the present system are correspondingly analogous to the coherent fields applied to drive the relevant atomic dipole-transitions.

As the typical application of quantum-optical analogy, we now analytically solve the coupled-mode equation (7) by generalizing the dynamical invariant method demonstrated by the Eqs.~(2-3) for analytically solving the corresponding time-dependent Schr\"odinger equation (1). To this end, corresponding the "Hamiltonian" (8) we first look for an invariant $I(z)$, which satisfies the condition:
\begin{align}
\frac{\partial{I(z)}}{\partial z}=-j[H(z), I(z)].
\end{align}
Such an invariant can be formally constructed as~\cite{CX5}:
\begin{align}
I(z)=\frac{1}{2} \begin{bmatrix} 0&\cos\gamma\sin\theta&-j\sin\gamma\\
\cos\gamma\sin\theta& 0 &\cos\gamma\cos\theta\\j\sin\gamma&\cos\gamma\cos\theta&0\end{bmatrix},
\end{align}
and the relevant parameters are determined by
\begin{align}
\Omega_{12}(z)=\frac{d\theta}{dz}\cot\gamma\sin\theta-\cos\gamma\sin\theta,
\end{align}
and
\begin{align}
\Omega_{23}(z)=\frac{d\theta}{dz}\cot\gamma\cos\theta-\cos\gamma\cos\theta,
\end{align}
\begin{align}
\Omega_{13}(z)=j\sin\gamma.
\end{align}
Here, $\gamma$ is a $z-$independent real parameter which is adjustable for various designs.
The eigenvectors of the invariant $I(z)$ can be easily written as:
\begin{align}
|\Phi_{0}(z)\rangle=\begin{bmatrix}\cos\gamma\cos\theta\\-j\sin\gamma\\
-\cos\gamma\sin\theta \end{bmatrix},
\end{align}
with the eigenvalue $0$, and
\begin{align}
|\Phi_{\pm}(z)\rangle=\frac{1}{\sqrt{2}}\begin{bmatrix}\sin\gamma\cos\theta \pm j\sin\theta\\j\cos\gamma\\
-\sin\gamma\sin\theta \pm j\cos\theta \end{bmatrix},
\end{align}
corresponding to the eigenvalues $\pm 1$, respectively. Similarly to construct the generic solution (3) of the Schr\"odinger equation (1) by using the eigenstates of the dynamical invariant $I(t)$, now the generic solution to the coupled-mode equation (7) can be expressed as
\begin{align}
|\Psi(z)\rangle=\sum_{m=0,\pm}C_m e^{j\alpha_m(z)}|\Phi_m(z)\rangle,
\end{align}
with
\begin{align}
\alpha_{m}(z)=\int_{0}^{z}\langle\Phi_{m}(z^{'})|j\frac{\partial}{\partial z'}-H(z')|\Phi_{m}(z')\rangle dz',
\end{align}
being the relevant Lewis-Riesenfeld phase. Specifically,
\begin{align}
\alpha_{0}(z)=0,
\end{align}
for the eigenstate $|\Phi_{0}(z)\rangle$, and
\begin{align}
\alpha_{\pm}(z)=\mp[\frac{\theta(z)-\theta(0)}{\sin\gamma}-z]
\end{align}
for the eigenstates $|\Phi_{\pm}(z)\rangle$.
Consequently, the modal amplitude $A_{m}(z)$ and light power $P_{m}(z)$ in the $m-$waveguide can be generically expressed as
\begin{align}
A_{m}(z)=\langle\Psi_{m}|\Psi(z)\rangle,\,P_{m}(z)=|A_{m}(z)|^2.
\end{align}
With the exact solution to the coupled-mode equation demonstrated above, the propagations of the electromagnetic waves along the three-waveguide coupled structure can be determined completely, in principle. Inversely, the three-waveguide devices for various applications can be designed precisely, once the relevant parameters $\theta$ and $\gamma$ are properly set. Note that the above solution to the coupled-mode equation is beyond the adiabatic limit, the three-wavguide coupled devices designed by using such an exact solution is compacted, which is useful to implement the device integrations.

\section{Designs of the 3D light couplers for the desired power divisions}
Given the exact solution to the coupled-mode equations constructed above, we are now discuss how to design the compacted 3D light couplers for the potential applications in integrated optics. Specifically, for a given initial condition, e.g., the signal inputting along certain waveguides, it is required to design the proper lengths of the waveguides and also the inter-waveguide coupling parameters for getting the desired outputs. Simply,
for a given input, e.g., $A_k(0)\neq 0$, we investigate how to design the proper the optimized parameters $\theta$ and $\gamma$ for realizing the stationary power outputs:
\begin{align}
P_m(L)=|A_{m}(L)|^2=|\langle\Psi_{m}|\Psi(L)\rangle|^2,\,m=1,2,3,
\end{align}
with as short as possible length $L$ of the waveguide device.
\begin{figure}[!t]
	\centering
	\includegraphics[width=8.3cm,height=11.7cm]{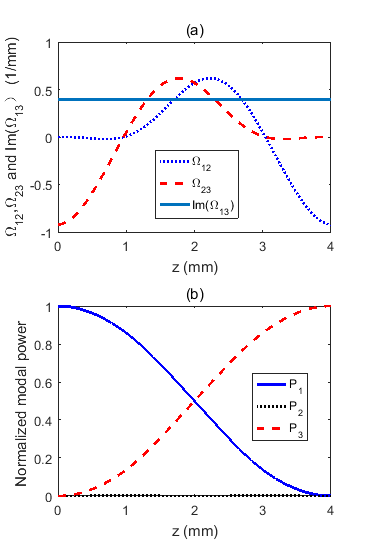}
		\caption{(a) $z$-dependent coupling coefficients $\Omega_{12}$, $\Omega_{23}$ and the imaginary part of $\Omega_{13}$ designed for implementing the given mode power conversion $P_{1}\rightarrow P_{3}$. (b) Evolution of the modal power in the three-waveguide coupler along $z$. Here, the relevant parameters in Eqs.~(26-28) are set as: $\varepsilon=0.4036, \eta=\pi/2, L=4mm$.}
\label{fig_2}
\end{figure}

Typically, we discuss how the power of the electromagnetic wave inputting along the waveguide $1$ is divided into the three waveguides throughout the 3D light coupler. In this case, the boundary conditions at the input port of the coupler reads $|\Psi(0)\rangle=|\Psi_{1}\rangle$ with $\gamma(0)=\varepsilon, \theta(0)=0$ and also
\begin{align}
C_{0}=\cos\varepsilon,\,C_{\pm}=\frac{1}{\sqrt{2}}\sin\varepsilon,
\end{align}
and
\begin{align}
\alpha_{0}(L)=0,\,\alpha_{\pm}(L)=\mp(\frac{\eta}{\sin\varepsilon}-L)=\mp\xi.
\end{align}
If the parameters $\theta(z)$ and $\gamma$ are designed as:
\begin{align}
\gamma=\varepsilon=const.,
\end{align}
and
\begin{align}
\theta(z)=\frac{10\eta}{L^{3}}z^{3}-\frac{15\eta}{L^{4}}z^{4}+\frac{6\eta}{L^{5}}z^{5},
\end{align}
with $\eta$ being a $z-$independent adjustable variable, then
the modal powers at the outputs could be designed as:
\begin{align}
P_{1}(L)=|A_{1}(L)|^{2}=&|\cos\eta[\cos^{2}\varepsilon+\sin^{2}\varepsilon\cos\xi],\nonumber \\
&+\sin\eta\sin\varepsilon\sin\xi|^{2}
\end{align}
\begin{align}
P_{2}(L)=|A_{2}(L)|^{2}=|j\cos\varepsilon\sin\varepsilon(\cos\xi-1)|^{2},
\end{align}
and
\begin{align}
P_{3}(L)=|A_{3}(L)|^{2}=&|-\sin\eta[\cos^{2}\varepsilon+\sin^{2}\varepsilon\cos\xi]\nonumber \\
&+\cos\eta\sin\varepsilon\sin\xi|^{2}.
\end{align}
with $\gamma(L)=\varepsilon, \theta(L)=\eta$. Therefore, once the relevant parameters have been properly designed, the desired 3D light couplers for various power divisions can be implemented.

\subsection{Modal power conversions}
 As the first example, we discuss how to design the 3D light couplers for power conversions, which require completely transfer the light power from one waveguide to another.
 Suppose that the light is initially putted along the waveguide 1 at $z=0$, with the modal function $|\Psi_{1}(0)\rangle$  and the power distribution $P(0)=(P_1(0),P_2(0),P_3(0))=(1, 0, 0)$.

In order to implement the modal power conversion, e.g.,$P_{1}\rightarrow P_{3}$, i.e., the light power distributions of the coupler at the output port should be $P(L)=(P_1(L),P_2(L),P_3(L))=(0, 0, 1)$, the parameters $\varepsilon$ and $\eta$ in Eqs.~(26-28) can be set as
\begin{align}
 \eta=\frac{\pi}{2},
 \end{align}
and
\begin{align}
\xi=\frac{\eta}{\sin\varepsilon}-L=2k\pi, (k=0,1,2,\cdots).
 \end{align}
Note that $k=0$ for $L\geq \pi/2$, and if $k$ is a non-zero integer the waveguide length $L$ can be arbitrarily short.
Specifically, for $k=0$ one can set
 \begin{align}
 \varepsilon=0.4036,\eta=\frac{\pi}{2}, L=4mm.
 \end{align}
 Substituting Eq.~(31) in Eqs.~(24-25), the parameters $\gamma$ and $\theta$ can be obtained as
\begin{align}
\gamma=\varepsilon=0.4036,
\end{align}
and
\begin{align}
\theta(z)=\frac{5\pi}{4^{3}}z^{3}-\frac{7.5\pi}{4^{4}}z^{4}+\frac{3\pi}{4^{5}}z^{5}.
\end{align}
respectively. With Eqs.~(12-14), once the coupling coefficients $\Omega_{12}(z)$ , $\Omega_{23}(z)$ and the imaginary part of $\Omega_{13}(z)$ are designed as those shown in Fig.~2(a),
Fig.~2(b) indicates clearly that the desired mode power conversion, i.e., $P_{1}(0)$ is perfectly converted into the $P_{3}(L)$.

Certainly, the length $L$ of the coupler can be further shorten by designing the proper parameters. For example, if we set
\begin{align}
 \varepsilon=\frac{\pi}{4},\eta=-1.5146, L=1\,mm,
 \end{align}
and also
\begin{align}
\gamma(z)=\varepsilon=\frac{\pi}{4},
\end{align}
\begin{align}
\theta(z)=-15.146z^{3}+22.719z^{4}-9.0876z^{5},
\end{align}
for the coupling coefficients $\Omega_{12}(z)$ , $\Omega_{23}(z)$ and the imaginary part of $\Omega_{13}(z)$ shown in Fig.~3(a), another modal power conversion: $P_{1}\rightarrow P_{2}$, can be
implemented:
\begin{align}
P_{1}(L)=0, P_{2}(L)=1, P_{3}(L)=0.
\end{align}
Here, the initial and final mode amplitude of the coupler are $\vec{A}(0)=(A_{1}(0),A_{2}(0),A_{3}(0))=(1,0,0)$ and $\vec{A}(L)=(A_{1}(L),A_{2}(L),A_{3}(L))=(0,-j,0)$, respectively. This means that the desired modal power conversion, i.e., $P_{1}(0)$ is perfectly converted into the $P_{2}(L)$. It is seen from Fig.~3(b) that, with such a short coupler one can robustly implement the modal power conversion, although the other modal could be populated temporarily during the power conversion.
\begin{figure}[!t]
	\centering
	\includegraphics[width=8.3cm,height=10.2cm]{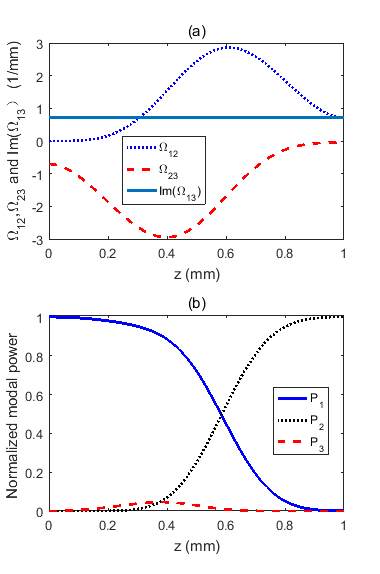}
	\caption{(a) $z$-dependent coupling coefficients $\Omega_{12}$, $\Omega_{23}$ and the imaginary part of $\Omega_{13}$ designed for implementing the given mode power conversion $P_{1}\rightarrow P_{2}$. (b) Evolution of the modal power in the three-waveguide coupler along $z$. The relevant parameters in Eqs.~(26-28) are set as follows: $\varepsilon=\pi/4, \eta=-1.5146, L=1mm$.}
	\label{fig_3}
\end{figure}
\subsection{Modal power divisions}
Below, we discuss how to design the 3D light power dividers, i.e., to implement the divisions of the optical power from the input waveguide to the other waveguides at the designable  splitting ratios.

First, let us consider how to implement the One-to-Two power divisions, i.e., the power in the waveguide 1 is averagely divided into the two waveguides.

(1) Case 1:\,$P_{1}\rightarrow \frac{1}{2}P_{2}+\frac{1}{2}P_{3}$

The light is initially in the waveguide 1 only, i.e., the modal function of the coupler at the input port is $|\Psi(0)\rangle=|\Psi_{1}\rangle$, and the modal powers at the output should be
\begin{align}
P_{1}(L)=0, P_{2}(L)=\frac{1}{2}, P_{3}(L)=\frac{1}{2}.
\end{align}
Using Eqs.~(26-28), we set
\begin{align}
 \varepsilon=\frac{\pi}{8}, \eta=\frac{\pi}{2}, L=0.9631mm
\end{align}
and also
\begin{align}
\gamma=\varepsilon=\frac{\pi}{8},
\end{align}
\begin{align}
\theta(z)=\frac{5\pi}{0.9631^{3}}z^{3}-\frac{7.5\pi}{0.9631^{4}}z^{4}+\frac{3\pi}{0.9631^{5}}z^{5}.
\end{align}
If the real coupling coefficients $\Omega_{12}$, $\Omega_{23}$ and the imaginary part of $\Omega_{13}$ are designed as those schematically shown in Fig.~4(a), then changes of the modal power distributions along the coupler can be implemented, see Fig.~4(b). After the power division, the modal function at the output port reads $|\Psi(L)\rangle=-j|\Psi_{2}\rangle/\sqrt{2}-|\Psi_3\rangle/\sqrt{2}$.
\begin{figure}[!t]
	\centering
	\includegraphics[width=8.8cm,height=11.7cm]{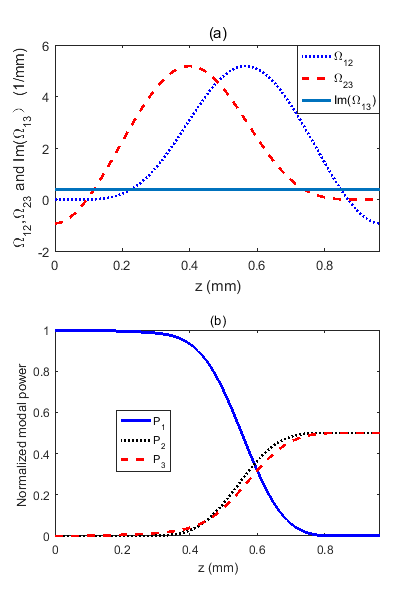}
	\caption{(a) $z$-dependent coupling coefficients $\Omega_{12}$, $\Omega_{23}$ and the imaginary part of $\Omega_{13}$ designed for implementing the given mode power splitting $P_{1}\rightarrow \frac{1}{2}P_{2}+\frac{1}{2}P_{3}$. (b) Evolution of the modal power in the three-waveguide coupler along $z$. The relevant parameters in Eqs.~(26-28) are set as follows: $\varepsilon=\pi/8, \eta=\pi/2, L=0.9631mm$.}
	\label{fig_4}
\end{figure}

(2) Case 2: $P_{1}\rightarrow \frac{1}{2}P_{1}+\frac{1}{2}P_{3}$

This means that the power of the light entering waveguide 1 is equally splitted into the  waveguides 1 and 3, i.e.,
\begin{align}
P_{1}(L)=\frac{1}{2}, P_{2}(L)=0, P_{3}(L)=\frac{1}{2}.
\end{align}
To this aim, we set
\begin{align}
 \varepsilon=0.031, \eta=\frac{\pi}{4}, L=0.2068mm
\end{align}
and
\begin{align}
\gamma=\varepsilon=0.031,
\end{align}
\begin{align}
\theta(z)=\frac{2.5\pi}{0.2068^{3}}z^{3}-\frac{3.75\pi}{0.2068^{4}}z^{4}+\frac{1.5\pi}{0.2068^{5}}z^{5},
\end{align}
Similarly, the real coupling coefficients $\Omega_{12}$, $\Omega_{23}$ and the imaginary part $\Omega_{13}$ are designed as those schematically shown in Fig.~5(a). It is seen from Fig.~5(b) that such a power division can be implemented, with the modal function at the outport:
$|\Psi(L)\rangle=|\Psi_1\rangle/\sqrt{2}-|\Psi_3\rangle/\sqrt{2}$.
\begin{figure}[!t]
	\centering
	\includegraphics[width=8.5cm,height=11.7cm]{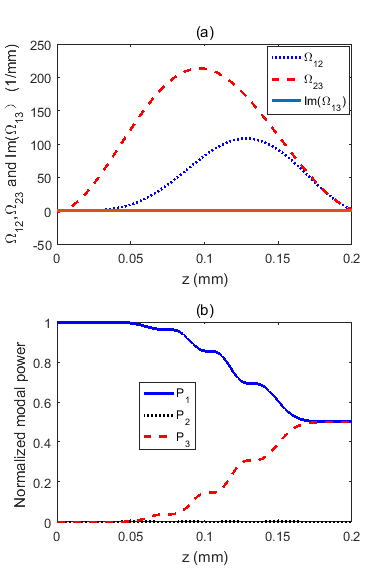}
	\caption{(a) $z$-dependent coupling coefficients $\Omega_{12}$, $\Omega_{23}$ and the imaginary part of $\Omega_{13}$ designed for implementing the given mode power splitting $P_{1}\rightarrow \frac{1}{4}P_{2}+\frac{3}{4}P_{3}$. (b) Evolutions of the modal powers in the three-waveguide coupler along $z$. The relevant parameters are set as: $\varepsilon=0.031, \eta=\pi/4, L=0.2068mm$.}
	\label{fig_5}
\end{figure}

It is emphasized that, the designs of the above parameters for various desired power transfers are not unique. For one function, many combinations of the parameters are possible. As a consequence, the length $L$ of the coupler is adjustable for the convenience of the fabrication.

Second, the power division with various splitting ratios can also be designed conveniently. For example,

(3) Case 3: $P_{1}\rightarrow \frac{1}{4}P_{2}+\frac{3}{4}P_{3}$.

To realize such an unequal power divisions, with
$P(0)=(P_{1}(0), P_{2}(0), P_{3}(0))=(1,0,0)$ and
the modal power at the output is
\begin{align}
P_{1}(L)=0, P_{2}(L)=\frac{1}{4}, P_{3}(L)=\frac{3}{4},
\end{align}
one may set
\begin{align}
 \varepsilon=\frac{\pi}{12}, \eta=\frac{\pi}{2}, L=2.93mm
\end{align}
and
\begin{align}
\gamma=\varepsilon=\frac{\pi}{12},
\end{align}
also
\begin{align}
\theta(z)=\frac{5\pi}{2.93^{3}}z^{3}-\frac{7.5\pi}{2.93^{4}}z^{4}+\frac{3\pi}{2.93^{5}}z^{5}.
\end{align}
As a consequence, if the coupling coefficients $\Omega_{12}$, $\Omega_{23}$ and the imaginary part of $\Omega_{13}$ are designed as those schematically shown in Fig.~6(a), then Fig.~6(b) shows that the desired power division could be implemented. Here, the modal function at the output port reads $|\Psi(L)\rangle=-j|\Psi_2\rangle/2-\sqrt{3}|\Psi_3\rangle/2$.
\begin{figure}[!t]
	\centering
	\includegraphics[width=8.5cm,height=11.7cm]{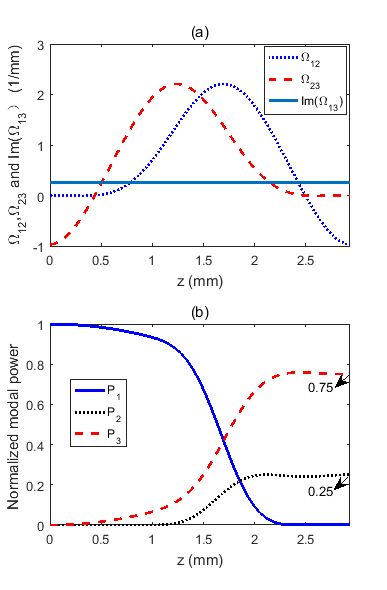}
	\caption{(a) $z$-dependent coupling coefficients $\Omega_{12}$ , $\Omega_{23}$ and the imaginary part of $\Omega_{13}$ designed for implementing the given mode power splitting $P_{1}\rightarrow \frac{1}{4}P_{2}+\frac{3}{4}P_{3}$. (b) Evolutions of the modal powers in the three-waveguide coupler along $z$. The relevant parameters are set as: $\varepsilon=\pi/12, \eta=\pi/2, L=2.93mm$.}
	\label{fig_6}
\end{figure}

Thirdly, the coupler could be designed as an One-to-Three power divider. For example,

(4) Case 4: $P_{1}\rightarrow \frac{1}{3}P_{1}+\frac{1}{3}P_{2}+\frac{1}{3}P_{3}$.

This requires that the power of the light entering along waveguide 1 is equally splitted into the waveguides 1, 2 and 3. The power distributions of the light at the outport of the coupler should be
\begin{align}
P_{1}(L)=\frac{1}{3}, P_{2}(L)=\frac{1}{3}, P_{3}(L)=\frac{1}{3}.
\end{align}
To this end, let us set
\begin{align}
 \varepsilon=0.3078, \eta=-\frac{\pi}{4}, L=0.5495mm
\end{align}
and
\begin{align}
\gamma=\varepsilon=0.3078,
\end{align}
\begin{align}
\theta(z)=\frac{2.5\pi}{0.5495^{3}}z^{3}-\frac{3.75\pi}{0.5495^{4}}z^{4}+\frac{1.5\pi}
{0.5495^{5}}z^{5}.
\end{align}
\begin{figure}[!t]
	\centering
	\includegraphics[width=8.3cm,height=11.7cm]{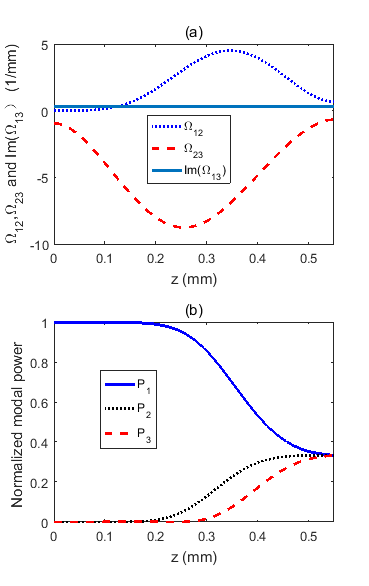}
	\caption{(a) $z$-dependent coupling coefficients $\Omega_{12}$ , $\Omega_{23}$ and the imaginary part of $\Omega_{13}$ designed for implementing the given mode power splitting $P_{1}\rightarrow \frac{1}{3}P_{1}+\frac{1}{3}P_{2}+\frac{1}{3}P_{3}$. (b) Evolution of the modal power in the three-waveguide coupler along $z$. The relevant parameters in Eqs.~(26-28) are set as follows: $\varepsilon=0.3078, \eta=-\pi/4,  L=0.5495mm $. }
	\label{fig_7}
\end{figure}
Then, we design the coupling coefficients $\Omega_{12}$, $\Omega_{23}$ and $\Omega_{13}$ as those schematically shown in Fig.~7(a). By numerical method, we see from the Fig.~7(b) that the desired power divisions at the output ports of the coupler can be implemented. Correspondingly, the modal function at the outports of the coupler reads:  $|\Psi(L)\rangle=|\Psi_1\rangle/\sqrt{3}-j|\Psi_2\rangle/\sqrt{3}+|\Psi_3\rangle/\sqrt{3}$.

Besides the designs demonstrated above, in general, this technique can be applied to deliver various 3D light couplers with arbitrary power ratios, once the relevant physical parameters, typically, e.g., $\gamma$ and $\theta(z)$, are designed properly.

\section{Discussion and Conclusion}
In the above designs and numerical simulations, the relevant parameters were usually assumed to be set precisely. In practical and also for the manufactures, various imperfections exist actually. Therefore, it is necessary to consider their robustness of the designs versus the parameter perturbations of the device, e.g., the geometric defects. In fact, the errors of the designs and manufactures may inevitably lead to the function imperfections, affecting the conversion efficiency and the accuracy of the split ratios of the light power. Therefore, the robustness of the designs is required to be examined, being directed against the imperfections of the set parameters.

Specifically, let us check the robustness of the designs against the fluctuations of the coupling coefficients, which exist usually in the fabrications. We introduce a parameter called the fidelity $F$ to describe the error robustness, e.g., $F_2$ and $F_3$ for the modal conversions:  $|\Psi_{1}\rangle \rightarrow  |\Psi_{2}\rangle$ and $|\Psi_{1}\rangle \rightarrow  |\Psi_{3}\rangle$, respectively. The fluctuations of the inter-waveguide coupling coefficients could be simply set as $\Omega_{12}(1+\delta)$, $\Omega_{13}(1+\delta)$ and $\Omega_{23}(1+\delta)$, with the relatively-high error range: $-0.1\leq \delta\leq 0.1$.
In Fig.~8 and Fig.~9, we shows how the fidelities: $F_{2}$ and $F_{3}$ of the power conversion devices change with the relatively error, respectively. One can see that the fidelities of the desired power  conversions are still high; the lowest fidelities are $F_{2,3}>97\%$. Therefore, the functions of the designed couplers are not very sensitive to the deviations of the coupling coefficients. This implies that the designs are robustness against the potential fabrication errors of the devices.
\begin{figure}[!t]
	\centering
	\includegraphics[width=7.2cm,height=4cm]{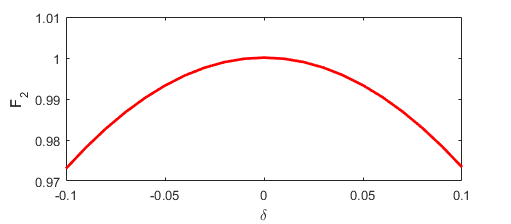}
	\caption{Fidelity $F_{2}$ of the power conversion: $|\Psi_{1}\rangle \rightarrow  |\Psi_{2}\rangle$, against the relative error of the coupling coefficients.}
	\label{fig_8}
\end{figure}

\begin{figure}[!t]
	\centering
	\includegraphics[width=7.2cm,height=4cm]{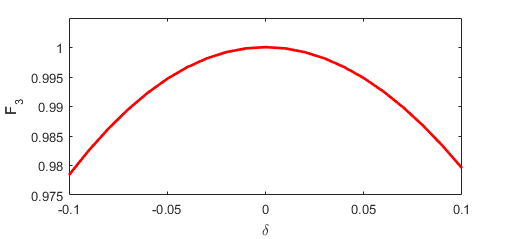}
	\caption{Fidelity $F_{3}$ of the power conversion: $|\Psi_{1}\rangle \rightarrow  |\Psi_{3}\rangle$, against the relative error of the coupling coefficient.}
	\label{fig_9}
\end{figure}

It is value to emphasize again that the refractive indexes in the real dielectric waveguides can be complex. For example, The refractive index of the multilayer planar waveguide structure can be expressed as~\cite{KHS,ENE}:
\begin{align}
N=N_{r}+jN_{i},
\end{align}
where $N_{r}>0$ and $N_{i}<0$. As a consequence, the coupling coefficient between the TE modes of the slab waveguides can be calculated as:
\begin{align}
\Omega_{ml}=\frac{k^{2}_{0}\int^{\infty}_{-\infty}\Psi^{*}_{m}(N^{2}-N^{2}_{m})\Psi_{l}dx}{2\beta_{m}\int^{\infty}_{-\infty}\Psi^{*}_{m}\Psi_{m}dx},
\end{align}
with $N_m$ is the refractive index profile of the planar waveguide structure, $\beta_{m}$ is the propagation constant and $\Psi_{m}$ the transverse mode field pattern of waveguid $m$. $k_{0}=2\pi/\lambda_{0}$ is the free-space wavenumber with $\lambda_{0}$ being the free-space wavelength. Therefore, our designs using the appropriate complex coupling coefficient $\Omega_{13}$ is feasible.

In summary, with the quantum-optical analogies we develop a systematic method to design the 3D light couplers for implementing various desired power divisions. The method is directly generalized from the time-domain dynamical invariant method, for exactly solving the Schr\"odinger equation of the driven quantum system, into the spatial-domain invariant for exactly solving the corresponding couple-mode equation of the three-waveguide coupler. With such an exact solution, the light power transfers between the waveguides in the coupler can be precisely designs by engineering the physical parameters of the coupler. Specifically, we demonstrated how to design various 3D couplers compactly for various desired power divisions. The fidelity of the designs against the errors of the typical coupling coefficients is also discussed. As the designs are beyond the usual adiabatic limit, the sizes of the devices could be as short as possible. This is particularly important for the 3D integrated optical technology.\\

{\bf Acknowledge.} This work is partly supported by the NSFC, Nos. 11974290 and
U1330201.

\ifCLASSOPTIONcaptionsoff
 \newpage
\fi

\end{document}